\documentclass[aps,preprint,preprintnumbers,nofootinbib,showpacs,superscriptaddress]{revtex4-1}
\usepackage{graphicx,color,amsmath}
\begin{document}
\title{Study of $\bar B\to \Lambda\bar p\,\rho(\phi)$ and $\bar B\to \Lambda\bar \Lambda \bar K^{*}$}

\author{C.Q. Geng}
\affiliation{Department of Physics, National Tsing Hua University, Hsinchu, Taiwan 300, R.O.C.}
\affiliation{Physics Division, National Center for Theoretical Sciences, Hsinchu, Taiwan 300, R.O.C.}
\author{Y.K. Hsiao}
\affiliation{Physics Division, National Center for Theoretical Sciences, Hsinchu, Taiwan 300, R.O.C.}
\date{\today}
\begin{abstract}
We examine the three-body baryonic $\bar B\to {\bf B\bar B'}V$ decays with $V$ vector mesons in the standard model.
We can simultaneously explain the recent experimental data on the decays of
$B^-\to \Lambda\bar p \,\rho^0$ and $\bar B\to \Lambda\bar \Lambda \bar K^{*}$
based on the QCD counting rules and $SU(3)$ flavor symmetry.
We also predict that the decay branching ratios of  $\bar B^0\to \Lambda\bar p \,\rho^+$ and
$B^-\to \Lambda\bar p \,\phi$ are $3.0$ and $1.5\times 10^{-6}$, respectively,
which are promising to be observed by BELLE and BABAR at  the $B$ factories.

\end{abstract}

\pacs{}

\maketitle
\newpage
\section{introduction}

Large decay branching ratios as well as  angular distribution asymmetries
have been measured in the charmless three-body baryonic $\bar B\to {\bf B\bar B'}M$ decays,
where $M$ can be either pesudo-scalar ($P$) or  vector ($V$) mesons
\cite{ppKstar_BABAR,ppKstar_BELLE,ppK_BABAR,ppKpi_BELLE,Lambdappi_BABAR,LambdapX_BELLE,
LambdaLambdaK_BELLE,Lambdapbarrho_BELLE}.
In particular,
the threshold effect of the sharp peak around the threshold area
in the dibaryon invariant mass spectrum
has been observed as a common feature
in these baryonic decays~\cite{ppKstar_BABAR,ppKstar_BELLE,ppK_BABAR,ppKpi_BELLE,Lambdappi_BABAR,LambdapX_BELLE,
LambdaLambdaK_BELLE,Lambdapbarrho_BELLE}, which
makes the decay branching ratios reach to the magnitude of order $10^{-6}$,
accessible to the detection at the  $B$ factories.
Furthermore,
it partly gives the answer of the non-observations of the two-body baryonic $\bar B$ decays \cite{HouSoni},
such as 
$\bar B^0\to p\bar p, \Lambda\bar \Lambda$, and $B^-\to \Lambda\bar p$  \cite{2body},
since the dibaryon of the two-body decays can only be created at the $m_{\bar B}$ scale without an extra meson to release the energy,
which is far from the enhancement of the dibaryon threshold area.

In this note, we consider the baryonic decays of $\bar B\to {\bf B\bar B'}V$, which have not been fully analyzed theoretically,
 unlike the corresponding pesudoscalar modes of $\bar B\to {\bf B\bar B'}P$. 
 %
Currently,  the decay branching ratios of the vector modes have been measured to be
\cite{ppKstar_BABAR,ppKstar_BELLE,LambdapX_BELLE,LambdaLambdaK_BELLE,Lambdapbarrho_BELLE}
\begin{eqnarray}\label{mdata}
{\cal B}(B^-\to p\bar p K^{*-})
&=& (5.3\pm 1.5\pm 1.3)\times 10^{-6}\;\text{(BABAR)}\,,\nonumber\\
&=& (3.38^{+0.73}_{-0.60}\pm 0.39)\times 10^{-6}\;\text{(BELLE)}\,,\nonumber\\
{\cal B}(\bar B^0\to p\bar p \bar K^{*0})
&=& (1.47\pm 0.45\pm 0.40)\times 10^{-6}\;\text{(BABAR)}\,,\nonumber\\
&=& (1.18^{+0.29}_{-0.25}\pm 0.11)\times 10^{-6}\;\text{(BELLE)}\,,\nonumber\\
{\cal B}(B^-\to \Lambda\bar \Lambda K^{*-})&=& (2.19^{+1.13}_{-0.88}\pm 0.33)\times 10^{-6}\;\text{(BELLE)}\,,\nonumber\\
{\cal B}(\bar B^0\to \Lambda\bar \Lambda \bar K^{*0})&=& (2.46^{+0.87}_{-0.72}\pm 0.34)\times 10^{-6}\;\text{(BELLE)}\,,\nonumber\\
{\cal B}(B^-\to \Lambda\bar p \,\rho^0)
&=& (4.78^{+0.67}_{-0.64}\pm 0.60)\times 10^{-6}\;\text{(BELLE)}\,,
\end{eqnarray}
at the  $B$ factories, respectively.
It is interesting to note that
$B^-\to p\bar p K^{*-}$ is among the first predicted baryonic modes in Ref. \cite{HY1}
with the predicted decay branching ratio  around $2\times 10^{-6}$.
The relation of ${\cal B}(B^-\to p\bar p K^{*-})>{\cal B}(\bar B^0\to p\bar p \bar K^{*0})$
can be traced to the deviation between
the  transition form factors of $B^-\to p\bar p$ and $\bar B^0\to p\bar p$
in the pQCD counting rules \cite{GengHsiao1}.
In addition,  $CP$ violation for $ B^-\to p\bar p K^{*-}$ \cite{GengHsiao2,GengHsiao2-1} is found to be
nearly 20\% of the world average \cite{ppKstar_BABAR,ppKstar_BELLE}
even though it is still inconclusive experimentally due to
the data errors.
%
 %
It is clear that   a
theoretical examination on $\bar B\to {\bf B\bar B'}V$ is needed
in order to fit  the experimental data.
In this paper, we shall develop a systematic study of the $\bar B\to {\bf B\bar B'}V$ decays
to not only  explain the measured data but also explore some other vector modes to see if they are also accessible to the $B$ factories.

The paper is organized as follows.
In  Sec. II, we provide the formalism, in which
we show the decay amplitudes of $\bar B\to \Lambda\bar p \,\rho$,
$B^-\to \Lambda\bar p \,\phi$, and $\bar B\to \Lambda\bar \Lambda \bar K^{*}$.
We also parameterize the matrix elements and
define the timelike and $\bar B\to {\bf B\bar B'}$ transition form factors.
The numerical results are performed  in Sec. III.
In Sec. IV, we present the conclusions.

\section{Formalism}
In terms of the effective Hamiltonian \cite{Hamiltonian,ali}
at  quark level, the decay amplitudes of $\bar B\to \Lambda\bar p \,\rho$ and $B^-\to \Lambda\bar p \,\phi$
can be factorized as
\begin{eqnarray}\label{amp1}
{\cal A_C}(\bar B\to\Lambda \bar p \,\rho)&=&\frac{G_F}{\sqrt 2}\bigg\{ (V_{ub}V_{us}^* a_1-V_{tb}V_{ts}^*a_4)
\langle \Lambda \bar p|(\bar s u)_{V-A}|0\rangle \langle \rho|(\bar u b)_{V-A}|\bar B\rangle\nonumber\\
&& +V_{tb}V_{ts}^*2a_6\langle \Lambda \bar p|(\bar s u)_{S+P}|0\rangle\langle \rho|(\bar u b)_{S-P}|\bar B\rangle\bigg\}\;,\nonumber\\
{\cal A_T}(B^-\to \Lambda\bar p \,\phi)
&=&\frac{G_F}{\sqrt 2}\bigg\{-V_{tb}V_{ts}^*\bigg [
(a_3+a_4-\frac{a_9}{2})\langle \phi|(\bar s s)_{V-A}|0\rangle
+a_5\langle \phi|(\bar s s)_{V+A}|0\rangle \bigg]\nonumber\\
&&\langle \Lambda\bar p|(\bar s b)_{V-A}|B^- \rangle
\bigg\}\;,
\end{eqnarray}
with the Fermi constant $G_F$, the CKM matrix elements $V_{q_iq_j}$, and the various currents as
$(\bar{q}_iq_j)_{V\pm A}=\bar{q}_i\gamma_\mu(1\pm\gamma_5)q_j$ and $(\bar{q}_iq_j)_{S\pm P}=\bar{q}_i(1\pm\gamma_5)q_j$.
The subscripts of $\cal C$ and $\cal T$ in Eq. (\ref{amp1}) denote the quark current and 
the $\bar B$ transition, respectively,
responsible for the creation of a baryon pair in the final states.
For $\bar B\to \Lambda\bar \Lambda \bar K^{*}$,
its decay amplitude is separated into two parts,
\begin{eqnarray}\label{amp2-1}
{\cal A}(\bar B\to \Lambda\bar \Lambda \bar K^{*})&=&
{\cal A_C}(\bar B\to\Lambda\bar \Lambda \bar K^{*})+{\cal A_T}(\bar B\to\Lambda\bar \Lambda \bar K^{*})\;,
\end{eqnarray}
where
\begin{eqnarray}\label{amp2-2}
{\cal A_C}(\bar B\to \Lambda\bar \Lambda \bar K^{*})&=& \frac{G_F}{\sqrt 2}\bigg\{\bigg
[V_{ub}V_{us}^* a_2\langle \Lambda\bar \Lambda|(\bar u u)_{V-A}|0\rangle\nonumber\\
&&-V_{tb}V_{ts}^*\bigg (a_3\langle \Lambda\bar \Lambda|(\bar u u+\bar d d+\bar s s)_{V-A}|0\rangle\nonumber\\
&&+a_4\langle \Lambda\bar \Lambda|(\bar s s)_{V-A}|0\rangle+a_5\langle \Lambda\bar \Lambda|(\bar u u+\bar d d+\bar s s)_{V+A}|0\rangle
\nonumber\\
&&+\frac{a_9}{2}\langle \Lambda\bar \Lambda|(2\bar u u-\bar dd-\bar ss)_{V-A}|0\rangle\bigg)\bigg]
\langle \bar K^*|(\bar s b)_{V-A}|\bar B\rangle
\nonumber\\
&&+V_{tb}V_{ts}^*2a_6\langle \Lambda\bar \Lambda|(\bar s s)_{S+P}|0\rangle \langle \bar K^*|(\bar s b)_{S-P}|\bar B\rangle\bigg\}\;,\nonumber\\
{\cal A_T}(\bar B\to \Lambda\bar \Lambda \bar K^{*})&=&\frac{G_F}{\sqrt 2}\alpha_{\bar K^*}\langle \bar K^*|(\bar s q')_{V-A}|0\rangle
\langle \Lambda\bar \Lambda|(\bar q' b)_{V-A}|\bar B\rangle\;,
\end{eqnarray}
with $q'=u(d)$,
corresponding to $B^-\to \Lambda\bar \Lambda K^{*-}$ ($\bar B^0\to \Lambda\bar \Lambda \bar K^{*0}$),
and
\begin{eqnarray}\label{amp2-3}
\alpha_{K^{*-}}&=&V_{ub}V_{us}^* a_1-V_{tb}V_{ts}^*a_4\,,\nonumber\\
\alpha_{\bar K^{*0}}&=&-V_{tb}V_{ts}^*a_4\,.
\end{eqnarray}
In Eqs. (\ref{amp1})-(\ref{amp2-3}),
the parameters $a_i$ are given by $a_i=c^{eff}_i+c^{eff}_{i\pm 1}/N_c$ for $i=$odd (even)
with the color number $N_c=3$, where we adopt the values of
the effective Wilson coefficients $c^{eff}_i$ from Refs. \cite{Hamiltonian,ali}.

To evaluate the decay branching ratios, one requires to know the form factors and decay constants
of the matrix elements in  Eq.~(\ref{amp2-2}).
In ${\cal A_C}(\bar B\to\Lambda \bar p \,\rho)$ and ${\cal A_C}(\bar B\to \Lambda\bar \Lambda \bar K^{*})$,
the form factors of the $\bar B\to \rho$ and $\bar K^*$ transitions
are given to incorporate with the timelike baryonic form factors
in the matrix elements of $0\to {\bf B\bar B'}$ with ${\bf B\bar B'}=\Lambda \bar p$ and $\Lambda\bar \Lambda$.
The matrix elements of the $\bar B\to V$ transition are written as \cite{BSW}
\begin{eqnarray}
\label{BtoV}
\langle V|(q_3 b)_{S-P}|\bar B\rangle
&=&2i\frac{m_V}{m_b}A_0\;\varepsilon^\ast\cdot p_{\bar B}\;,\nonumber\\
\langle V|(q_3 b)_V|\bar B\rangle&=&\epsilon_{\mu\nu\alpha\beta}
\varepsilon^{\ast\nu}p_{\bar B}^{\alpha}p_{V}^{\beta}\frac{2V_1}{m_{\bar B}+m_{V}}\;,\nonumber\\
\langle V|(q_3 b)_A|\bar B\rangle
&=&i\bigg[\varepsilon^\ast_\mu-\frac{\varepsilon^\ast\cdot
q}{t}q_\mu\bigg](m_{\bar B}+m_{V})A_1
 + i\frac{\varepsilon^\ast\cdot q}{t}q_\mu(2m_{V})A_0\nonumber\\
&-&i\bigg[(p_{\bar B}+p_{V})_\mu-\frac{m^2_{\bar B}-m^2_{V}}{t}q_\mu \bigg](\varepsilon^\ast\cdot q)\frac{A_2}{m_{\bar B}+m_{V}}\;,
\end{eqnarray}
where $t\equiv q^2$ with $q=p_{\bar B}-p_{V}=p_{\bf B}+p_{\bf \bar B'}$,
$V_1$ and $A_{0,1,2}$ are the form factors, and
$\varepsilon^\ast_\mu$ is the polarization of $V=\rho\,(K^*)$ with $q_3=u\,(s)$, respectively.
For the dibaryon creations, we write
\begin{eqnarray}\label{form3}
\langle {\bf B}{\bf\bar B'}|(\bar q_1 q_2)_V|0\rangle
&=&\bar u(p_{\bf B})\bigg\{F_1(t)\gamma_\mu+\frac{F_2(t)}{m_{\bf B}
+m_{\bf \bar B'}}i\sigma_{\mu\nu}(p_{\bf \bar B'}+p_{\bf B})_\mu\bigg\}v(p_{\bf \bar B'})
\nonumber\\
&=& \bar u(p_{\bf B})\bigg\{[F_1(t)+F_2(t)]\gamma_\mu+\frac{F_2(t)}{m_{\bf B}+m_{\bf \bar B'}}
(p_{\bf \bar B'}-p_{\bf B})_\mu\bigg\}v(p_{\bf \bar B'})\;,\nonumber\\
\langle {\bf B}{\bf\bar B'}|(\bar q_1 q_2)_A|0\rangle&=&
\bar u(p_{\bf B})\bigg\{g_A(t)\gamma_\mu+\frac{h_A(t)}{m_{\bf B}+m_{\bf \bar
B'}} (p_{\bf \bar B'}+p_{\bf B})_\mu\bigg\}\gamma_5 v(p_{\bf \bar B'})\,,\nonumber\\
\langle {\bf B}{\bf\bar B'}|(\bar q_1 q_2)_S|0\rangle &=&f_S(t)\bar u(p_{\bf B})v(p_{\bf\bar B'})\;,\nonumber\\
 \langle {\bf B}{\bf\bar B'}|(\bar q_1 q_2)_P|0\rangle &=&g_P (t)\bar u(p_{\bf B})\gamma_5 v(p_{\bf\bar B'})\,,
\end{eqnarray}
where ${\bf B\bar B'}=\Lambda \bar p$ with $\bar q_1 q_2=\bar s u$,
${\bf B\bar B'}=\Lambda\bar \Lambda$ with $\bar q_1 q_2=\bar u u,\,\bar d d$, or $\bar s s$,
and $F_1$, $F_2$, $g_A$, $h_A$, $f_S$ and $g_P$ are the form factors.
In terms of the pQCD counting rules \cite{Brodsky1,Brodsky2,Brodsky3}, the momentum dependences of
the form factors $F_1(t)$ and $g_A(t)$ behave as $1/t^2$,
which characterizes two hard gluon exchanges between the valence quarks.
Explicitly, one obtains
\begin{eqnarray}\label{timelikeF2}
F_1(t)=\frac{C_{F_1}}{t^2}\bigg[\text{ln}\bigg(\frac{t}{\Lambda_0^2}\bigg)\bigg]^{-\gamma}\;, \qquad g_A(t)=\frac{C_{g_A}}{t^2}\bigg[\text{ln}\bigg(\frac{t}{\Lambda_0^2}\bigg)\bigg]^{-\gamma}\;,
\end{eqnarray}
where $\gamma=2+4/(3\beta)=2.148$ with $\beta$ being the QCD $\beta$ function and $\Lambda_0=0.3$ GeV.
Under the $SU(3)$ flavor and $SU(2)$ spin symmetries,
 the constants $C_{F_1}$ and $C_{g_A}$ can be related by another set of constants, given by \cite{GengHsiaoHY}
\begin{eqnarray}\label{C||}
&&C_{F_1}=C_{g_A}=-\sqrt\frac{3}{2}C_{||}\,, \;\;\;
\;\;\;\;\;\;\;\;\;\;\;\;\;\;\;\;\;\;\;\;\;\;\;\;\;\text{for $\langle \Lambda\bar p|(\bar s u)_{V,A}|0\rangle$},\nonumber\\
&&C_{F_1}=C_{g_A}=C_{||}\,,\;\;\;\;\;\;\;\;\;\;\;\;
\;\;\;\;\;\;\;\;\;\;\;\;\;\;\;\;\;\;\;\;\;\;\;\;\;\text{for $\langle \Lambda\bar \Lambda|(\bar s s)_{V,A}|0\rangle$},\nonumber\\
&&C_{F_1}=\frac{1}{2}C_{||}+\frac{1}{2}C_{\overline{||}}\,,\;
C_{g_A}=\frac{1}{2}C_{||}-\frac{1}{2}C_{\overline{||}}\,,\;\;\; \text{for $\langle \Lambda\bar \Lambda|(\bar q_i q_i)_{V,A}|0\rangle$},\;
\end{eqnarray}
with $q_i=u$ and $d$, respectively. Similarly, one has~\cite{ChuaHou}
\begin{eqnarray}\label{gp}
g_P=f_S\;.
\end{eqnarray}
By adding
\begin{eqnarray}\label{fsha}
f_S(t)&=&\frac{m_{\bf B}-m_{\bf B'}}{m_{q_1}-m_{q_2}}F_1(t)\;,
\nonumber\\
h_A(t)&=&-\frac{(m_{\bf B}+m_{\bf B'})^2}{t}g_A(t)\;,
\end{eqnarray}
which are derived in  equation of motion,
all the timelike baryonic form factors can be connected. 
With the inputs of the values of $C_{||}$ and $C_{\overline{||}}$ in Eq. (\ref{C||}),
we are able to estimate the values of the other timelike form factors by the relations in Eqs. (\ref{timelikeF2})-(\ref{fsha}).
Note that $F_2$ is suppressed by $1/(t\text{ln}[t/\Lambda_0^2])$ in comparison with $F_1$ \cite{F2,F2b}
and  can be safely ignored in the discussion.

 For ${\cal A_T}(B^-\to \Lambda\bar p \,\phi)$ and ${\cal A_T}(\bar B\to \Lambda\bar \Lambda \bar K^*)$,
the $0\to V$ ($V=\phi\,,\bar K^*$) transitions
can be parameterized as
\begin{eqnarray}\label{dc}
\langle V|\bar q_1 \gamma_\mu q_2|0\rangle&=&m_V f_{V}\varepsilon_\mu^*\,,
\end{eqnarray}
with the decay constants $f_{V}$.
The most general forms of
the $\bar B\to {\bf B\bar B'}$ transitions are expressed as \cite{AngdisppK}
\begin{eqnarray}\label{transitionF}
&&\langle {\bf B\bar B'}|(\bar q_3 b)_V|\bar B\rangle=
i\bar u(p_{\bf B})[  g_1\gamma_{\mu}+g_2i\sigma_{\mu\nu}p^\nu +g_3p_{\mu} +g_4q_\mu +g_5(p_{\bf\bar B'}-p_{\bf B})_\mu]\gamma_5v(p_{\bf \bar B'})\,,\nonumber\\
&&\langle {\bf B\bar B'}|(\bar q_3 b)_A|\bar B\rangle=
i\bar u(p_{\bf B})[ f_1\gamma_{\mu}+f_2i\sigma_{\mu\nu}p^\nu +f_3p_{\mu} +f_4q_\mu +f_5(p_{\bf\bar B'}-p_{\bf B})_\mu]        v(p_{\bf \bar B'})\,,
\end{eqnarray}
with $p=p_{\bar B}-p_{\bf B}-p_{\bf\bar B'}$, and $f_i$ and $g_i$ the form factors.
The momentum dependences of $f_i$ and $g_i$ are represented as \cite{ChuaHouTsai,GengHsiao3}
\begin{eqnarray}\label{figi}
f_i=\frac{D_{f_i}}{t^n}\,,\;\;g_i=\frac{D_{g_i}}{t^n}\,,
\end{eqnarray}
with $n=3$, where $D_{f_i}$ and $D_{g_i}$ are the constants to be determined by the $\bar B\to p\bar p M$ data,
while the form factors of  $B^-\to \Lambda\bar p$, $\bar B\to\Lambda\bar \Lambda$,
and $\bar B\to p\bar p$ are in the same forms.
We note that the number $n=3$ is set for three hard gluons as the propagators to form a baryon pair
in the approach of the pQCD counting rules,
where two of them attach to valence quarks in $\bf B\bar B'$,
while the third one kicks and speeds up the spectator quark in $\bar B$.
Again, with  the $SU(3)$ flavor and $SU(2)$ spin symmetries,
$D_{g_i}$ and $D_{f_i}$ from the vector currents are related by another set of constants $D_{||}$ and  $D_{\overline{||}}$ of the chiral currents.
Explicitly, we have \cite{GengHsiaoHY,Hsiao}
\begin{eqnarray}\label{D||-Lambdapbar}
D_{g_1}=D_{f_1}=-\sqrt\frac{3}{2}D_{||}\,,\;D_{g_j}=-D_{f_j}=-\sqrt\frac{3}{2}D_{||}^j\,,
\end{eqnarray}
with $j=2,3, ..., 5$ for $\langle \Lambda\bar p|(\bar s b)_{V(A)}|B^-\rangle$, and
\begin{eqnarray}\label{D||}
D_{g_1}=\frac{1}{2}D_{||}-\frac{1}{2}D_{\overline{||}}\,,\;
D_{f_1}=\frac{1}{2}D_{||}+\frac{1}{2}D_{\overline{||}}\,,\;
D_{g_j}=D_{f_j}=0\,,
\end{eqnarray}
for $\langle \Lambda\bar \Lambda|(\bar d b)_{V(A)}|\bar B^0\rangle$
and $\langle \Lambda\bar \Lambda|(\bar u b)_{V(A)}|B^-\rangle$.

\section{Numerical analysis}

We now specify various input parameters for our numerical analysis.
The vector decay constants in Eq. (\ref{dc}) are given by \cite{decayconst}
\begin{eqnarray}
(f_\phi,\;f_{\bar K^*})=(0.231,\;0.217)\; \text{GeV}\,,
\end{eqnarray}
and the $\bar B \to V$ transition form factors in Eq. (\ref{BtoV}) as the functions of $t$ are written as \cite{MFD}
\begin{eqnarray}
V_{1}[A_{0}](t)&=&\frac{V_{1}[A_{0}](0)}{(1-t/M_{1[2]}^2)
(1-\sigma_1 t/M_{1[2]}^2+\sigma_2 t^2/M_{1[2]}^4)}\,,\nonumber\\
A_{1,2}(t)&=&\frac{A_{1,2}(0)}{1-\sigma_1 t/M_2^2+\sigma_2 t^2/M_2^4}\,,
\end{eqnarray}
where the values of $V_1(0)$, $A_{0,1,2}(0)$, $\sigma_{1,2}$ and $M_{1,2}$ are shown in
Table \ref{MF} \cite{MFD}.
\begin{table}[h!]
\caption{ \sl The form factors of $\bar B\to \bar K^{*}\,(\rho)$ at $t=0$
in Ref. \cite{MFD} with  $(M_1,M_2)=(5.37,5.42)$
and $(5.27,5.32)$ GeV for  $K^*$ and  $\rho$, respectively.}\label{MF}
\begin{tabular}{|c|c|c|c|c|}
\hline
$B\to K^{*}\,(\rho)$&$V_1$&$A_0$&$A_1$&$A_2$\\\hline
f(0)       &0.44 (0.31) &0.45 (0.30) &0.36 (0.26) &0.32 (0.24)\\
$\sigma_1$ &0.45 (0.59) &0.46 (0.54) &0.64 (0.73) &1.23 (1.40)\\
$\sigma_2$ &-----&-----&0.36 (0.10) &0.38 (0.50)\\\hline
\end{tabular}
\end{table}

Based on the approach in Ref. \cite{NF_GengHsiao},
we refit the baryonic form factors
with the measured branching ratios of $\bar B^0\to n\bar p D^{*+}$, $\bar B^0\to \Lambda\bar p \pi^+$
and $B^-\to \Lambda\bar \Lambda K^{-}$, and  obtain
\begin{eqnarray}\label{C1}
C_{||}=136.4\pm 8.2\,\text{GeV}^{4}\;,\;\;
C_{\overline{||}}=13.8\pm 127.0\,\text{GeV}^{4}\;.
\end{eqnarray}
Note that since ${\cal A_T}(B^-\to \Lambda\bar p\, \phi)$,
${\cal A_T}(\bar B\to \Lambda\bar \Lambda \bar K^{*})$, and
${\cal A_T}(\bar B\to \Lambda\bar \Lambda \bar K)$ \cite{GengHsiao5}
are in association with
the $\bar B\to p\bar p M$ decays with $M=D^{(*)},\,\bar K^{*}$, and $\pi$, we have
adopted the values of the $\bar B\to {\bf B\bar B'}$ transition form factors
in Refs. \cite{GengHsiaoHY,Hsiao,GengHsiao6,GengHsiao4},
 given by
\begin{eqnarray}\label{inputD}
&&(D_{||},\;D_{\overline{||}})=(67.7\pm 16.3,\,-280.0\pm 35.9)\;{\rm GeV^5},\nonumber\\
&&(D_{||}^2,\,D_{||}^3,\,D_{||}^4,\,D_{||}^5)=\nonumber\\
&&(-187.3\pm 26.6,\,-840.1\pm 132.1,\,-10.1\pm 10.8,\,-157.0\pm 27.1)\;{\rm GeV^4}\;.
\end{eqnarray}
The squared amplitudes by summing over baryon spins denoted as $|\bar {\cal A}|^2$
now become available, of which the integration over the phase space in three-body decays is required.
Explicitly, by using
\begin{eqnarray}
d\Gamma=\frac{1}{(2\pi)^3}\frac{|\bar {\cal A}|^2}{32M^3_B}dm^2_{\bf B\bar B'}dm^2_{{\bf \bar B'}{V}}\,,
\end{eqnarray}
with $m^2_{{\bf B\bar B'}}=(P_{\bf B}+P_{\bf \bar B'})^2$ and $m^2_{{\bf \bar B'}{V}}=(P_{\bf \bar B'}+P_{V})^2$,
we get
 the numerical results for the branching ratios,
 summarized in Table \ref{tab1}.
\begin{table}[t!]
\caption{Numerical results in units of $10^{-6}$ for the branching ratios are presented in column 2,
where the theoretical errors come from the uncertainties in  the form factors,
in comparison with the world averaged experimental data in column 3.}\label{tab1}
\begin{tabular}{|c|c|c|}
\hline
decay mode & our result&  data\\\hline
$B^-\to \Lambda\bar p \,\rho^0$                &$3.28\pm 0.31$ &$4.78\pm0.90$\\
$\bar B^0\to \Lambda\bar p \,\rho^+$           &$3.01\pm 0.31$ &---\\
$B^-\to \Lambda\bar p \,\phi$                  &$1.51\pm 0.28$ &---\\
$B^-\to \Lambda\bar \Lambda K^{*-}$            &$1.91\pm 0.20$ &$2.19\pm 1.18$\\
$\bar B^0\to \Lambda\bar \Lambda \bar K^{*0}$  &$1.76\pm 0.18$&$2.46\pm 0.93$\\
\hline
\end{tabular}
\end{table}

It is worth to point out that the measured values of
${\cal B}(\bar B^-\to \Lambda\bar p \,\rho^0)$ and ${\cal B}(\bar B\to \Lambda\bar \Lambda \bar K^{*})$
are not involved in the fitting. 
As seen from Table \ref{tab1},
our result of ${\cal B}(B^-\to \Lambda\bar p \,\rho^0)=3.28\times 10^{-6}$ 
is now able to explain the data. This is mainly due to the relation of $g_P=f_S$ in Eq. (\ref{gp})
to fully take into account $g_P$,
whereas the use of equation of motion with the chiral behavior
may underestimate $g_P$ \cite{ChuaHou}.
It is interesting to note that
the decay of  $\bar B^0\to \Lambda\bar p \,\pi^+$ is similar to its corresponding vector mode,
which was  measured to be ten times bigger than the previous calculation \cite{HY1,ChuaHouTsai}.
The contributions of the scalar and pseudo-scalar currents are thus compatible,
and the $a_6$ term in $\cal A_C$ of Eq. (\ref{amp1})  gives the main contribution due to the
chiral enhancement.
In fact, 
 $\bar B\to \Lambda\bar \Lambda \bar K^{*}$
is also dominated by the $a_6$ term in $\cal A_C$ 
since the contribution of $g_2(f_2)\sigma_{\mu\nu}q^\nu$ 
prevails over those of the other terms in the $\bar B\to {\bf B\bar B'}$ transition form factors
with $D_{g_2}=D_{f_2}=0$ in Eq. (\ref{D||}) as demonstrated in the study of $\bar B\to p\bar p \bar K^*$.
Apart from ${\cal B}(B^-\to p\bar p\rho^-)\sim 2\times 10^{-5}$~ \cite{GengHsiao1},
${\cal B}(\bar B^0\to \Lambda\bar p \,\rho^+)=3.01\times 10^{-6}$ and
${\cal B}(B^-\to \Lambda\bar p \,\phi)=1.51\times 10^{-6}$ can be
within reach with the data sample accumulated by BELLE and BABAR at the $B$ factories.

\section{Conclusions}

We have analyzed the $\bar B\to {\bf B\bar B'}V$ decays based on the approach of the pQCD counting rules and $SU(3)$ flavor and $SU(2)$ spin
symmetries. We have found that the decay branching ratio of $B^-\to \Lambda\bar p \,\rho^0$
 is ten times bigger than the previous theoretical prediction and agrees with the current experimental data.
We have also shown that
$\bar B\to \Lambda\bar \Lambda \bar K^{*}$ is dominated by
the $a_6$ term in $\cal A_C$, and its value is comparable to the measured data as well.
In addition, we predict
${\cal B}(\bar B^0\to \Lambda\bar p \,\rho^+)=3.0\times 10^{-6}$ and
${\cal B}(B^-\to \Lambda\bar p \,\phi)=1.5\times 10^{-6}$, which
are promising to be observed by  BELLE and BABAR as well as LHCb.

\begin{acknowledgments}
The work was supported in part by National Center of Theoretical Science
and  National Science Council
(NSC-98-2112-M-007-008-MY3)
of R.O.C.
\end{acknowledgments}

\end{document}